\documentclass[twocolumn,showpacs,preprintnumbers,amsmath,amssymb]{revtex4}
\usepackage{graphicx}

\newcommand{\ea}{{\em et al.~}}

\begin{document}

\title{Mechanism for nonequilibrium symmetry breaking and pattern
formation in magnetic films}

\author{J. M. Deutsch and Trieu Mai}
\affiliation{
Department of Physics, University of California, Santa Cruz, 
California 95064, USA}

\date{\today}

\begin{abstract}
Magnetic thin films exhibit a strong variation in properties depending
on their degree of disorder. Recent coherent x-ray speckle experiments
on magnetic films have measured the loss of correlation between
configurations at opposite fields and at the same field, upon
repeated field cycling. We perform finite temperature numerical
simulations on these systems that provide a comprehensive explanation
for the experimental results. The simulations demonstrate, in
accordance with experiments, that the memory of configurations
increases with film disorder.  We find that non-trivial microscopic
differences exist between the zero field spin configuration obtained
by starting from a large positive field and the zero field configuration
starting at a large negative field.  This seemingly paradoxical
behavior is due to the nature of the vector spin dynamics and is
also seen in the experiments.  For low disorder, there is an
instability which causes the spontaneous growth of line-like domains
at a critical field, also in accord with experiments.  It is this
unstable growth, which is highly sensitive to thermal noise, that
is responsible for the small correlation between patterns under
repeated cycling.  The domain patterns, hysteresis loops, and memory
properties of our simulated systems match remarkably well with the
real experimental systems.
\end{abstract}


\maketitle 

\section{Introduction} 
Many magnetic systems have a memory of their past configurations.
This history, manifest in the hysteresis loop of a magnet, has many
fascinating features, for example Barkhausen noise which demonstrates
that this history has intricate behavior at a small scale, with
avalanches occurring in the same order in a highly reproducible
manner~\cite{Bark, Barkhaus2, Barkhaus3, Barkhaus4, Barkhaus6,
Barkhaus7, Barkhaus8, Barkhaus9}.  From a theoretical perspective,
a certain class of Ising models has been proved by Sethna \ea
~\cite{Sethna} to exhibit perfect return point memory \footnote{Sethna's
definition of return point memory refers to systems that have the
property that  their spins will return to exactly the same configuration
after an excursion of the external field away from and then back
to its original value.  The excursion cannot exceed  the maximum
and minimum values that had already been taken by the field.} And
of course, hysteresis is the principle that makes magnetic storage
devices possible~\cite{techref}.

Primarily due to the technological importance of magnetic systems,
a vast number of experiments have been done to observe magnetic
memory and hysteresis phenomena~\cite{techref}.  A comparably large
number of theories have been made to explain these experiments with
varying degrees of success.  A recent experiment by Pierce \ea
~\cite{Sorensen,Sorensen2} measured the memory properties of magnetic
multilayer thin films.  They observed an effect that at first sight
appears paradoxical, involving fundamental issues of symmetry in
these systems.  In this paper, through numerical simulations, we
attempt to provide an explanation for this effect.

\begin{figure}
\begin{center}
\includegraphics[width=3in]{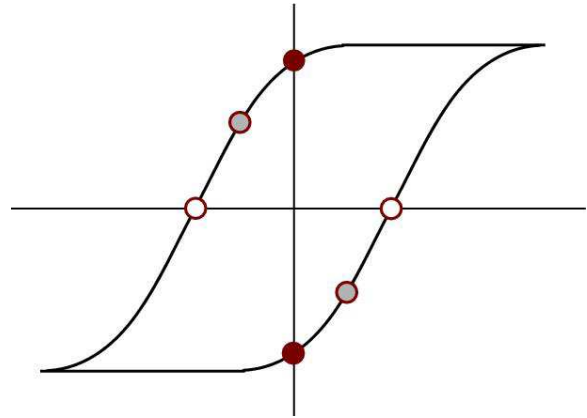}
\caption{Magnetization vs. external field for a cartoon hysteresis
system. Return point memory is defined as the covariance between
any point with the same point after an integer number of complete
cycles.  Complementary points are points lying a half-integer number
of field cycles away.  Each pair of points (filled, shaded, and
unfilled) is a pair of complementary points.}
\label{exhyst}
\end{center}
\end{figure}

The experiment by Pierce \ea used the powerful new tool of x-ray
speckle metrology to measure the covariance of a domain configuration
at one field with the configuration at another field.  This measured
value is one way to quantify the amount of ``memory'' possessed by
the material.  Measurements were done for Co/Pt multilayer samples
with varying amounts of disorder; the study found that samples with
greater disorder had higher ``memory'' than the ordered samples
which showed no significant amount of memory.  The domain patterns
ranged from labyrinthine mazes for low disorder samples to ones
without any noticeable structure for the highest disorder samples.
Hysteresis loops for different samples had dramatically different
features depending on the amount of disorder, with steep cliffs
(large changes in the magnetization at constant field) existing in
the hysteresis loops of the low disorder samples.

In addition to the memory dependence on disorder, an unexpected
finding was the difference between what Pierce \ea called ``return
point memory'' (RPM) and ``complementary point memory'' (CPM).  RPM
was defined by them as the covariance, see below, between the
configuration at a certain field point with the configuration at
the same field point after an integer number of complete cycles
around the major hysteresis loop.  CPM is defined as the covariance
between the configuration at a certain field point with the
configuration at a half integer number of field cycles away.  Figure
\ref{exhyst} shows examples of ``return'' and ``complementary''
points.  The covariance that we will use in order to define RPM and
CPM is
\begin{equation}
cov(a,b) = \langle {\bf s}_a({\bf r}) \cdot {\bf s}_b({\bf r}) 
\rangle - \langle {\bf s}_a({\bf r}) \rangle \cdot \langle 
{\bf s}_b({\bf r}) \rangle,
\label{cov}
\end{equation}
where $a$ and $b$ refer to legs on the hysteresis loop and the
average is over all space (spins).  The normalized covariance is
defined by
\begin{equation}
\rho = \frac{cov(a,b)}{\sqrt{cov(a,a) cov(b,b)}}.
\label{normcov}
\end{equation}
At a certain external field $B_e$, RPM is defined as $\rho(B_e,a;B_e,b)$,
where $a$ and $b$ are legs going in the same direction (both ascending
or both descending), and CPM is defined as $-\rho(B_e,a;-B_e,b)$,
where $a$ and $b$ are legs going in opposite directions.  Perfect
memory would lead to both RPM and CPM equal to unity.  A covariance
between two independent systems would approach zero for both RPM
and CPM.

The speckle experiment measured intensities from x-ray scattering.
Because scattering experiments probe samples in Fourier space, the
actual measurements did not measure the covariance as defined by
Eq.(\ref{cov}) but instead by
\begin{equation}
cov_k(a,b) = \langle |s^z_a({\bf k})|^2 |s^z_b({\bf k})|^2 \rangle - 
\langle |s^z_a({\bf k})|^2 \rangle \langle |s^z_b({\bf k})|^2 \rangle,
\label{covk}
\end{equation}
where $s^z_a({\bf k})$ is the z component of the Fourier transform
of the spins on leg $a$ and the the average is over wave vectors
\footnote{The actual experimentally measured quantity differs from
Eq. (\ref{covk}) slightly, because the average intensities depend
on $k$ and so the formula needs to be suitably weighted.
see~\cite{Sorensen,Sorensen2} for details.}.  Using this Fourier
space covariance in the definition of RPM and CPM, Pierce \ea
measured the amount of memory for samples of Co/Pt multilayer thin
films with a wide range of disorder.  RPM and CPM are both negligible
for samples with low disorder and are quite significant ($\gtrsim
0.6$ at the coercive field) for highly disordered samples.  The
onset of memory occurs quite abruptly as disorder is increased.
Furthermore, for samples with significant memory, RPM is noticeably
larger than CPM for all fields measured.  In this paper we will use
the real space covariance in the definition of RPM and CPM.

A viable model of the Co/Pt multilayer films must at the least be
able to explain two experimental results:
\begin{enumerate}
\label{item:rpm=0}
\item  RPM, CPM $\sim 0$ for systems with low disorder
\item  $1 >$ RPM $>$ CPM for systems with high disorder
\label{item:rpm>cpm}
\end{enumerate}
In this paper, we provide a possible explanation for these experimental
results through finite temperature numerical simulations of classical
vector spins.  The finite temperature destroys perfect memory and
affects both RPM and CPM. We will see that the nature of the domains
determine how much the temperature affects the covariance; highly
ordered systems are more susceptible to temperature effects than
highly disordered systems, and do explain result \ref{item:rpm=0}.
But thermal noise does not discriminate between the ascending leg
and the descending leg.  A satisfactory theory must also show that
RPM is greater than CPM.

Condition \ref{item:rpm>cpm} is difficult to understand intuitively.
Consider a system at low temperature that starts out fully saturated
from being in a large positive external field perpendicular to the
film, and then that field is brought down adiabatically to zero.
At remanence the system will be in a state with domains pointing
in different directions.  Now repeat the same procedure but starting
with a fully saturating field pointing in the opposite direction.
At remanence we now expect the final configuration to be the same
apart from a change in sign ${\bf s} \rightarrow -{\bf s}$. The
fact that the CPM value is less than the RPM value contradicts this.
The configurations although somewhat correlated, are different.

One possible explanation of this can be devised by introducing a
term in the Hamiltonian that is not invariant under spin and external
field reversal ${\bf s} \rightarrow -{\bf s}$ and $B_e \rightarrow
-B_e$.  Terms in this class automatically introduce a difference
between the ascending and descending legs.  A system with such a
Hamiltonian would exhibit a hysteresis loop that is not symmetric
under the same operations even at zero temperature.  Using a scalar
$\phi^4$ model with a random field term in the Hamiltonian,
Jagla~\cite{Sorensen2, Jagla} was able to satisfy both conditions.
But the physical origin, and existence, of these random fields is
unclear.  Recently, with some modifications to his $\phi^4$ model,
including replacing the random fields with random anisotropy,
Jagla~\cite{Jagla2} was able to produce domain patterns and hysteresis
loops that resemble the experimental patterns and loops remarkably
well.  However without a random field term in the Hamiltonian, the
memory conditions could not be satisfied.

We provide a possibly more fundamental mechanism that satisfies
condition \ref{item:rpm>cpm}: The vector dynamics breaks the spin
and field reversal symmetry, thereby reducing CPM and not RPM.  This
mechanism does not require any new terms in the Hamiltonian of the
class mentioned in the previous paragraph.  In our earlier
work~\cite{pillar}, we showed how the vector dynamics, governed by
the Landau-Lifshitz-Gilbert (LLG) equation~\cite{LLGref}, can give
rise to non-complementary hysteresis loops for a system of nanomagnetic
pillars.  The crucial role of vector dynamics for explaining
experimental results underlines the inadequacy of scalar models,
even when the system is highly anisotropic.  We have not excluded
the possibility that  the real explanation is a combination of the
vector dynamics and the random fields. Furthermore it is possible
that the experiments result from some other effects that have not
been considered so far. In this paper, we use numerical simulations
to demonstrate the plausibility of this vector mechanism.

In the next section of this paper, the LLG equation is introduced
and the various terms in the Hamiltonian and their origin will be
described.  Details of the numerics are provided in Sec. III and
the types of domain patterns and hysteresis loops from these
simulations are presented in Sec. IV.  Sec. V contains the covariance
results.

\section{Classical Spin Dynamics and the model Hamiltonian}
The Landau-Lifshitz-Gilbert equation of motion~\cite{LLGref}  is
the simplest dynamic equation for classical spins that contains a
reactive term and a dissipative term:
\begin{equation}
\frac{d{\bf s}}{dt} = -{\bf s} \times ({\bf B} 
-\gamma {\bf s} \times {\bf B}),
\label{LLG}
\end{equation}
where ${\bf s}$ is a microscopic classical spin, ${\bf B}$ is the
local effective field, and $\gamma$ is a damping coefficient.  The
effective field is ${\bf B} = - \partial\mathcal{H}/\partial{\bf
s}+\mbox{\boldmath$\zeta$}$, where $\mathcal{H}$ is the Hamiltonian
and $\mbox{\boldmath$\zeta$}$ represents the effect of thermal
noise.  The overall constant in front of the right hand side of
Eq.(\ref{LLG}) is set to unity.

Both terms on the right hand side of the LLG equation are necessary
to adequately describe the motion of classical spins.  The double
cross product in the damping term is necessary in order to maintain
a constant magnitude for the spins.  The relaxation time is inversely
related to the damping coefficient. The reactive term describes
precession of a spin about its local effective field and has a
quantum mechanical origin; the commutation relations of angular
momentum variables give rise to the single cross product of the
reactive term.  The commutation relation between spin variables is
odd under spin reversal.  Later, we will show how precession is
crucial to adequately describe the experimental system using our
Hamiltonian.  Because scalar models cannot have precession, the
mechanism that we suggest is not possible in scalar theories.

The LLG equation can be applied to microscopic spins as well as
coarse grained spins.  Every spin variable in the numerics represents
a block spin of the multilayer film and the evolution of all spins
is calculated by numerically integrating the LLG equation. Henceforth
in this paper, a ``spin'' corresponds to a block spin variable.
The coefficients of the Hamiltonian, the damping coefficient, and
thermal noise are also those associated with the coarse grained
variables.  The Hamiltonian has four terms: self anisotropy energy,
local ferromagnetic interactions, long range dipole-dipole interactions,
and the energy for the interactions with the external field.

The Co/Pt multilayer film is a perpendicular anisotropic material.
We define this perpendicular axis as the $z$ axis.  The origin of
the perpendicular easy axis is the layered structure of the material.
Any real material will have imperfections in the layering, and this
imperfection in the layering is the ``disorder'' mentioned many
times in the previous section.  Samples with low disorder show very
well defined planes separating the layers of the two elements,
whereas samples with high disorder have very rough interfaces between
the Co and Pt.  Due to this disorder in the layering, each spin has
a different easy axis, ${\bf \hat n}_i$. The anisotropy energy term
is described by
\begin{equation}
\mathcal{H}_{ani} = -\alpha \displaystyle\sum_{i} 
({\bf s}_{i} \cdot {\bf \hat n}_i)^2,
\label{ani}
\end{equation}
where $\alpha$ is a model parameter.  The anisotropy must be an
even function due to the symmetry in $\pm {\bf \hat n}_i$.  Aside
from being even, the functional form of Eq.(\ref{ani}) can be quite
complicated in general, but if a power expansion is possible,
Eq.(\ref{ani}) would be the leading order term.

The next term in the Hamiltonian describes the local ferromagnetic
coupling between neighboring spins,
\begin{equation}
\mathcal{H}_{J} = - \displaystyle J\sum_{<i,j>}  
{\bf s}_i \cdot {\bf s}_j,
\end{equation}
where $J$ is the ferromagnetic coupling constant.  We consider only
the $J>0$ case.  Even though we attempt to model a continuum system,
a grid is necessary for the numerics.  In order to minimize artificial
effects due to the grid, this real space ferromagnetic energy term
is replaced by a term in Fourier space of the form
\begin{equation}
\mathcal{H}_{el} = J \displaystyle \sum_{\bf k} k^2 s_{\bf k}^2.
\label{elastic}
\end{equation}
Eq.(\ref{elastic}) is the leading order term of the elastic energy
in Fourier space which aligns spins locally similar to $\mathcal{H}_J$.
Disorder can be introduced in the elastic constant $J$ as well.
For small disorder of this type, the results presented in this paper
do not change qualitatively. We eliminate this extra disorder
parameter and have a uniform elastic constant for all spins.

In addition to the local interactions, the spins also interact
via a long range dipole-dipole energy.  The form of this energy is
the usual classical expression
\begin{equation}
\mathcal{H}_{dip} = -w \displaystyle\sum_{i,j\neq i} 
\frac{3({\bf s}_i \cdot {\bf e}_{ij})({\bf e}_{ij} \cdot {\bf s}_j) 
- {\bf s}_i \cdot {\bf s}_j}{r_{ij}^3},
\label{dip}
\end{equation}
iwhere ${\bf r}_{ij}$ is the displacement vector between spins $i$
and $j$, ${\bf e}_{ij}$ is the unit vector along this direction,
and $w$ is another model parameter.  When the spins are pointing
predominantly in the $\pm z$ directions, the dipolar fields tend
to anti-align the spins.  This competition with the ferromagnetic
interaction produces many of the domain features.  Interestingly,
when the spins are in-plane, the dipole-dipole and local interactions
can become cooperative (depending on ${\bf r}_{ij}$).

The form of the dipole energy expressed by Eq.(\ref{dip}) is correct
for point dipoles, but must be corrected for the block spins of
interest.  The small but finite thickness of the layers does not
significantly change the long range behavior of Eq.(\ref{dip}), but
it does eliminate the divergence at small
separations~\cite{LangerGoldstein}.  This cutoff is implemented in
Fourier space which is a more computationally efficient basis to
calculate long range dipolar fields.  Multiplying a gaussian of the
form $\exp(-k^2d^2/2)$ to the Fourier transform of Eq.(\ref{dip})
effectively removes the divergence at large wave vectors while
retaining the original form for small wave vectors.  The model
parameter, $d$, is the length scale of the cutoff where the finite
thickness of the sample affects the dipolar interaction.

The last term in the Hamiltonian is the energy from the interaction
of the spins with the external field which we take to be uniform in
the $z$ direction,
\begin{equation}
\mathcal{H}_{ext} = -B_e \displaystyle\sum_i s_i^z.
\label{ext}
\end{equation}
Major hysteresis loops are simulated by cycling the external field
from a large positive value (past saturation) to a large negative
value and back again.  To ensure total saturation we reset the value
of the spin variables to point completely along the field direction
at the start of each leg in the hysteresis loop.

In summary, the full model Hamiltonian has four terms: anisotropy,
ferromagnetic coupling, dipole-dipole interaction, and the external
field term as described by Eqs.(\ref{ani},\ref{elastic}-\ref{ext}),
\begin{equation}
\mathcal{H} = \mathcal{H}_{ani}+\mathcal{H}_{el}+\mathcal{H}_{dip}
+\mathcal{H}_{ext}. 
\label{hamilt}
\end{equation}
The model Hamiltonian is bilinear in the spins, ${\bf s}_i$ and
external field, $B_e$, therefore it is symmetric under ${\bf s}_i
\rightarrow -{\bf s}_i$ and $B_e \rightarrow -B_e$.  However, as
mentioned above, this symmetry does not exist for the equation of
motion.  Under these two operations, the left hand side of the LLG
equation, Eq.(\ref{LLG}), and the dissipative term reverse signs,
whereas the reactive term stays unchanged.  With the symmetric model
Hamiltonian and using purely relaxational dynamics, in some sense
there would be no way to differentiate from going down the descending
leg or up the ascending leg of the hysteresis loop.  The difference
in sign change between the terms of the LLG equation under spin and
field reversal is the mechanism by which RPM $>$ CPM.  When precession
is turned off, the covariance results presented in Sec. V are no
longer valid and we verified that as expected, RPM $\simeq$ CPM
when the temperature is non-zero and RPM = CPM = 1 identically when
there is no thermal noise.

This explanation for why RPM $>$ CPM may become invalid when the
relaxation time is much smaller than the precessional period.  There
is no reason to believe that this is the case for multilayer thin
films.  In fact, $\gamma$, the damping coefficient which is a measure
of the relative importance between damping and precession has been
measured for NiFe thin films and found to be approximately
0.01~\cite{Sandler}.  This indicates that the relaxation time is
much larger than the precessional period, thus precession is
substantial for some systems.  The damping is found to be enhanced
to $\sim 1$ for CoCr/Pt multilayer films~\cite{Lyberatos}.  For the
numerical simulations, we have set $\gamma$ to unity.

\section{Numerics}
We simulate the multilayer thin film by a two dimensional lattice
of block vector spins of unit length.  Most of the model parameters
are noted in the previous section: the relative strengths of the
coupling constants in the Hamiltonian, $\alpha, J,$ and $w$, the
dipolar cutoff length $d$, and the damping coefficient $\gamma$.
There are two more model parameters, the temperature, $T$, and
$\lambda$, a parameter that controls the variation in the easy axes,
${\bf \hat n}_i$.  Each ${\bf \hat n}_i$ is assigned randomly  for
every spin.  The stacking of the layers in the $z$ direction is the
physical origin of the anisotropy, therefore the ${\bf \hat n}_i$'s
are weighted in this direction.  This weighting factor, $\lambda$,
is inversely related to the amount of disorder.

The large number of parameters may at first seem hopeless from the
point of view of prediction because in a seven dimensional parameter
space almost any curve can likely be fit. This clearly diminishes
the predictive power of a model.  All is not lost, because the
behavior described in this paper is quite robust in many of these
parameters. Furthermore there is agreement, at least at the qualitative
level, for all the quantities measured at a fixed value of parameters:
the shape of the hysteresis loops as a function of disorder, the
evolution at low disorder of patterns in the films which have also
been seen experimentally, and the RPM and CPM behavior as a function
of disorder. As we mentioned at the beginning of the last section,
$\gamma$ has been measured in a certain multilayer system to be
$\sim 1$~\cite{Lyberatos}, and we have set it equal to unity for
all simulations.  A smaller $\gamma$ would almost certainly enhance
our results.  The dipolar cutoff, $d$, determines a length scale
of the domains, and all other parameters are considered with respect
to this length scale.  Without loss of generality, we have set $d
= 4$ lattice spacings.  Below, we will report results for different
temperature and disorder, therefore $T$ and $\lambda$ are not fixed
model parameters.  That leaves us with a three dimensional parameter
space and the task does not seem so daunting as before.

One may ask that since we are simulating a known experimental system,
why are we not using measured values for these quantities in our
numerics? First, these are not known for these disordered Co/Pt
films, and even if this could be done, the simulations are of coarse
grained spins and not microscopic spins therefore all coefficients,
for example, temperature, magnetic field, disorder strength. would
change in a highly nontrivial manner.  Lattices of sizes $128\times
128$ and $256\times 256$ spins are used in the simulations, therefore
each block spin represents roughly forty thousand real microscopic
spins.  Nevertheless, even with only $128^2$ spins, we are able to
observe behavior quite similar to what is seen in the experiments.

Initially, the easy axes, ${\bf \hat n}_i$ are independently and
randomly assigned for each spin.  The $z$ component is weighted by
$\lambda$ which sets the amount of disorder; disorder is small when
$\lambda$ is large and is increased by decreasing $\lambda$.  The
system is initially saturated in the positive $z$ direction.  Starting
from a large positive external field, $B_e^{max}$, the field is
adiabatically decreased to a large negative field, $-B_e^{max}$ and
the spins are again saturated.  From there, the field is adiabatically
increased back to $B_e^{max}$ to finish one complete field cycle.

At zero temperature, adiabatic field cycling is straightforward:
after every field step, the spins evolve in time until the configuration
reaches a local minimum in the energy, then the field is changed
by another step.  With thermal noise, the adiabatic condition cannot
be as stringent as it is at absolute zero. The adiabatic condition
becomes nontrivial because thermal noise is constantly buffeting
the spins. The systems of interest have a large anisotropy with
easy axes weighted in the $z$ direction, therefore the energy minima
of the spins are predominantly in the $\pm z$-direction.  If the
temperature is not too large, it would be unlikely for thermal noise
to be sufficient to knock the spin over the energy barrier to induce
a spin-flip.

We use the fact that thermally activated spin-flips are uncommon
to construct an adiabatic condition.  The spins are allowed to vary
their orientation from the vertical, but if the $z$ component of
any spin changes from greater than a large threshold value to less
than the negative of that same value (or vice versa), then the
system has not yet satisfied the adiabatic condition.  If spin-flip
events do occur, the spins must be evolved further without changing
the field.  This threshold value is set to 0.75 for the results in
this paper.  Qualitatively similar results are obtained for slightly
different thresholds.  In addition to not allowing large spin flips
for the adiabatic condition, the change of the total magnetization
in the $z$ direction, $\Delta M$, must also not change substantially.
The condition used is $\Delta M < 0.1$.

The time evolution of the spins follows the LLG equation.  This is
done by calculating the effective field for each spin and then
integrating the equations using the fourth order Runge-Kutta
algorithm.  Due to the long range nature of the dipolar interaction,
all the spins are coupled together and a real space numerical
integration is $O(N^2)$.  Using a FFT algorithm, this is reduced
to $O(N \ln N)$.  Periodic boundary conditions are used to accommodate
the Fourier method, but the results here should not differ significantly
with different boundary conditions.

The configurations of spins are stored at certain fields and RPM
and CPM as defined in the first section are calculated between
different legs.  For efficiency, many legs are run in parallel.  In
the next section, the domain patterns and hysteresis loops resulting
from these numerics are discussed.

\section{Domain Patterns and Hysteresis Loops}
Using the numerics described in the previous section, qualitatively
different configurations and hysteresis loops arise depending on
the parameters described in Sec. II.  The important parameters are
the relative strengths between the interaction terms in the Hamiltonian
($\alpha, J$, and $w$), and the amount of disorder (characterized
by $\lambda$).  The coarse graining introduces difficulties in
calculating these parameters from the microscopic interactions.
Furthermore, even a first principles calculation of the microscopic
interactions is highly nontrivial.  For example, the anisotropy of
the spins is believed to be due to a quantum exchange mechanism
between layers that is not easily calculated or understood.
Nevertheless, we can proceed by empirically searching through
parameter space to find parameters in which the experimental results
are observed.

\begin{figure}
\begin{center}
\includegraphics[width=3.25in]{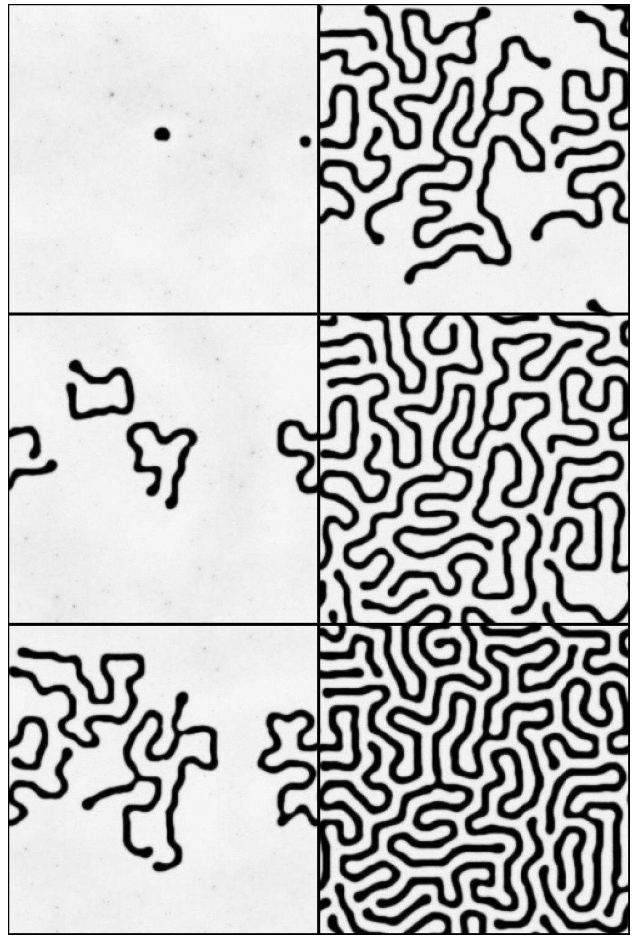}
\caption{Domain growth for a $256^2$ system with low disorder,
$\lambda=1000$ and $w=0.15$.  The $z$ component of the spins are
shown with light shades representing ``up'' and dark shades
representing ``down.'' The first five panels are at constant field
and show snake-like domains growing in time.  The labyrinthine
patterns resemble the domain patterns of the low disorder samples
from the speckle experiments.  Eventually, the length of the domains
becomes comparable to the system size and require a lowering of the
external field to increase in size.  The last panel is at a lower
external field than the other five.}
\label{linegrow}
\end{center}
\end{figure}

In the region of parameter space in which the domain patterns and
hysteresis loops have properties seen in the real experimental
samples, the ferromagnetic coupling $J$ and the anisotropy strength
$\alpha$ are approximately equal, whereas the dipole-dipole strength
$w$ is roughly an order of magnitude weaker.  The competitive nature
of these interactions give rise to interesting domain patterns and
hysteresis loops~\cite{Jagla, Jagla2}.  A strong $\alpha$ tries to
keep all spins pointing out of plane due to the weighting in the
$z$ direction.  The positive $J$ tries to maintain a local alignment
of spins, thus large domains are favored by this energy term.  Though
weaker in magnitude per spin-spin interaction, the dipolar force
is long range with an accumulated effect that competes significantly
with the local couplings.

The amount of disorder in a system is determined by $\lambda$, with
a large $\lambda$ signifying low disorder.  When $\lambda = 1$, the
anisotropy energy is spherically symmetric (after averaging over
many spins) and the system is no longer a perpendicular material.
In addition to lowering $\lambda$, disorder also weakens the dipolar
strength.  Disorder in the real experimental systems exists in the
interface between the Co and Pt layers.  When disorder is large,
the interface is highly non-planar, thus effectively suppressing
the dipolar interaction after coarse graining.  Though this disorder
in the interface would most likely alter the relative strengths of
$J$ and $\alpha$, we remove this degree of freedom and set $J =
0.85$ and $\alpha = 0.875$ for all systems.

The low disorder systems have the most interesting domain patterns
with labyrinthine mazes at remanence.  The systems with the lowest
amount of disorder studied have $\lambda = 1000$ and $w = 0.15$.
For this value of $\lambda$, the easy axes ${\bf \hat n}_i$ is
almost parallel to ${\bf \hat z}$.  Starting at positive saturation,
the large $J$ and $\alpha$ keep the spins aligned essentially
vertically.  But the dipole-dipole interaction is highly dissatisfied
in this configuration.  Because of this dissatisfaction, at a
relatively high positive field, one or more small patches will flip.

Once a small ``down'' domain has formed in a sea of up spins, the
local ferromagnetic interaction causes instabilities to occur at
the boundary of the domain.  Due to the ordered environment around
the domain, the instabilities grow into lines.  These configurations
are long serpentine configurations in accordance with the recent
experiments on Co/Pt films and also earlier experimental
work~\cite{SeulMonar}.  The growth of the snakes is spontaneous,
occurring at constant external field.  Figure \ref{linegrow} shows
the formation and  growth of the snakes.  In regions of high
curvature, occasionally side branches grow and look very similar
to experimental patterns. Due to the dipolar force, the domains
behave as if they repel each other.  This repulsion causes the
growth to halt once the length of the snakes is comparable to the
system size and the total area of negative spins fills a finite
fraction of the sample. At this point, in order for more spins to
flip, the field must be lowered.

The initial circular instability and subsequent growth occurs for
a wide range of temperatures and therefore appears to be predominantly
different from what one would expect if this was nucleation.  If
this was nucleation, near the critical field, domains would appear
and disappear with a temperature dependent probability.  Our
simulations show that when a domain flips initially, it remains in
that state above the critical field and never disappears.  At the
critical field a domain will quickly become unstable, even at zero
temperature, and grows into serpentine patterns even with no change
in applied field.  The initial instability often occurs at the same
location upon repeated field cycling which further indicates that
this is not nucleation.

The growth of these snake-like domains is similar to what is seen
in ferrofluids~\cite{LangerGoldstein,ferrofluid1,MirandaWidom}.  In
that case however volume of fluid is conserved which means that the
volume of fluid cannot spontaneously increase with external field.
However the general shape of domains is similar, particularly their
long serpentine shape ending with a wider head as shown by figure
\ref{linegrow}.

The growth at constant field introduces a cliff in the hysteresis
loop; the growth of the snakes reduces the magnetization abruptly.
Hysteresis curves of this type are shown in the first panel of
figure \ref{hysts}.  The end of this large avalanche signifies the
point at which the snakes have taken up the available growth
environment and are intertwined.  After this point, the length of
the snakes can no longer increase significantly but the width of
the snakes does gradually increase by lowering the external field.
Eventually, the field overcomes the repulsion and the snakes link
and form a labyrinthine maze.  Finally, the field will have a large
negative value that saturates the spin in the negative $z$ direction.

The systems with the highest amount of disorder studied have $\lambda
= 2$ and $w = 0.05$.  With such a low weighting, the easy axes have
large components in the plane that vary greatly from one spin to
the next.  Because the ${\bf \hat n}_i$'s no longer have a large
probability of pointing in the $z$ direction and the anisotropy
energy is strong, the saturation magnetization is reduced.  In other
words, the spins tend to point along their easy axes which have
larger in-plane components than the low disorder systems.  This
reduction of the saturation magnetization as disorder is increased
is also seen in the experiments.  Though the spins are more in-plane,
the external field still starts at a large positive value, therefore
the $z$ component of the spins start positive.  The weaker dipole
coefficient requires more reduction of the field before the dipolar
dissatisfaction causes a patch to flip, thus the remanent magnetization
increases for more disordered systems.

\begin{figure}
\begin{center}
\includegraphics[width=3.25in]{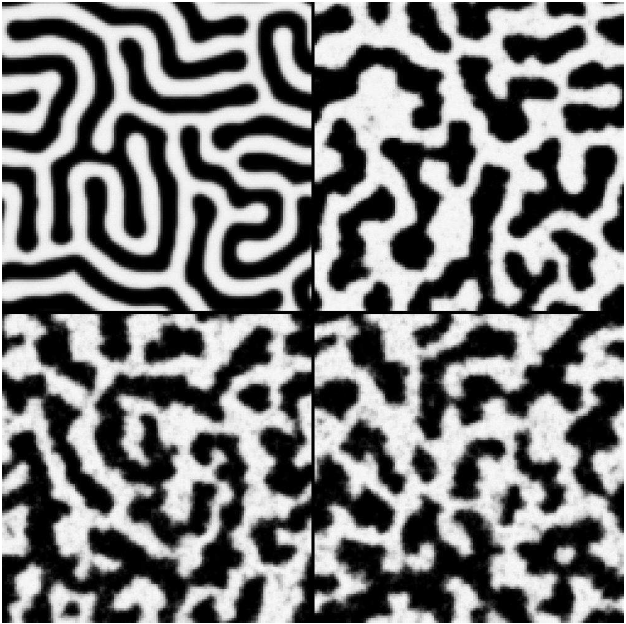}
\caption{Spin configurations at the coercive field for systems with
different amounts of disorder.  The systems with the least amount
of disorder are in the upper left panel and greatest amount are in
the lower right.  The disorder parameters $\lambda$ are $1000, 11,
4.1, 3$ and the dipole strengths $w$ are $0.15, 0.105, 0.08, 0.06$.
}
\label{disorder}
\end{center}
\end{figure}

\begin{figure}
\begin{center}
\includegraphics[width=1.65in]{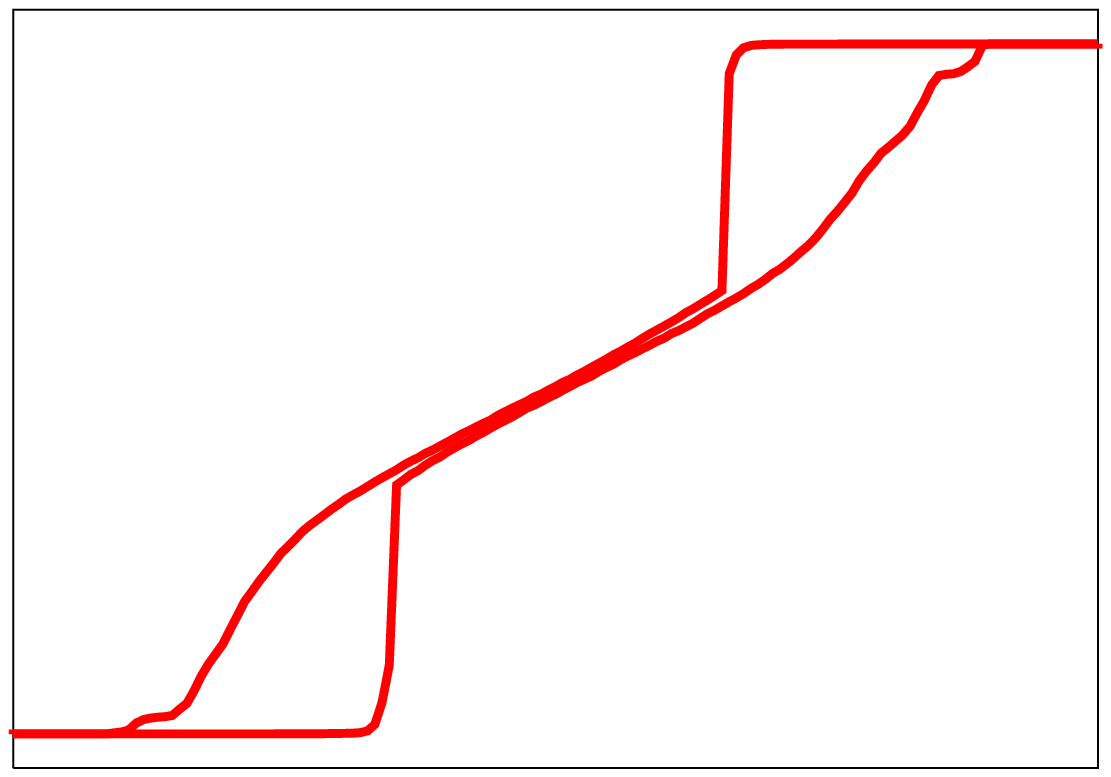}
\includegraphics[width=1.65in]{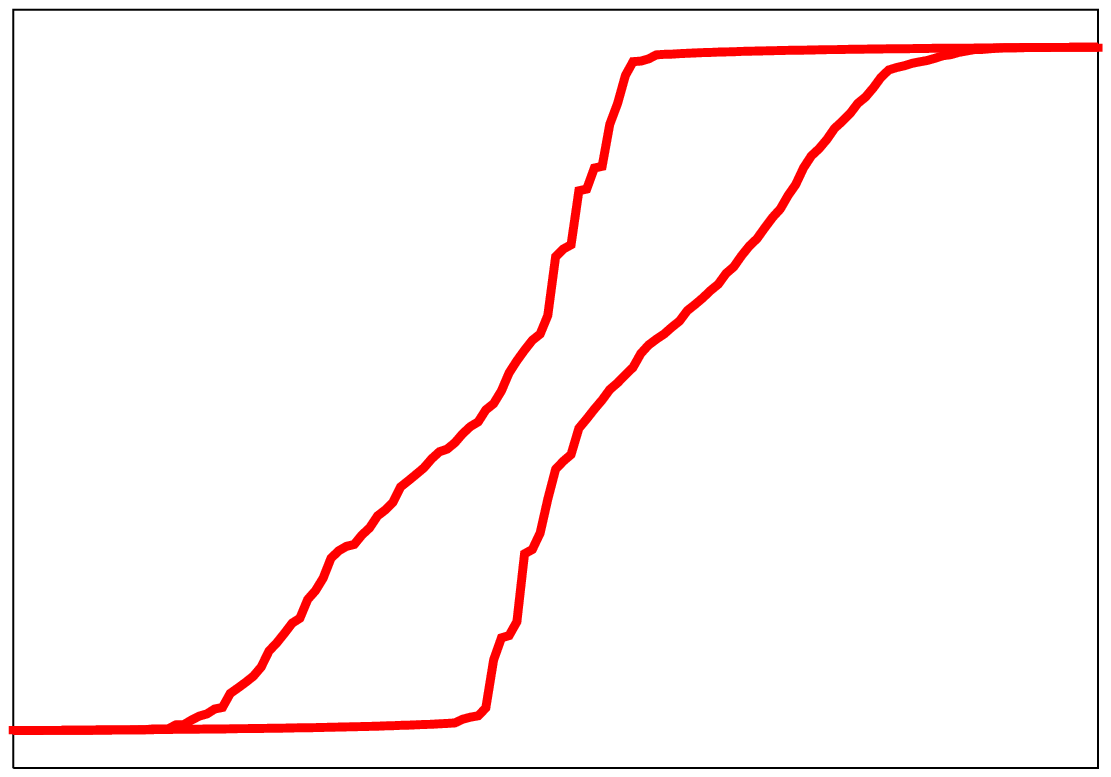}\\
\includegraphics[width=1.65in]{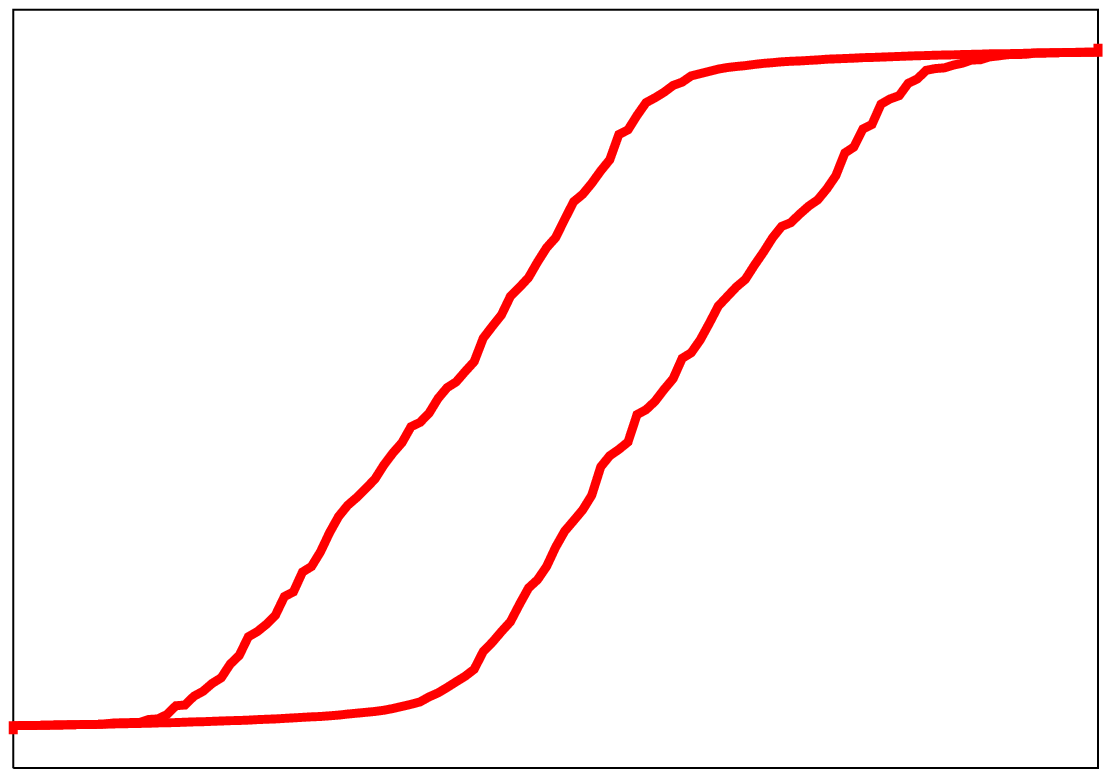}
\includegraphics[width=1.65in]{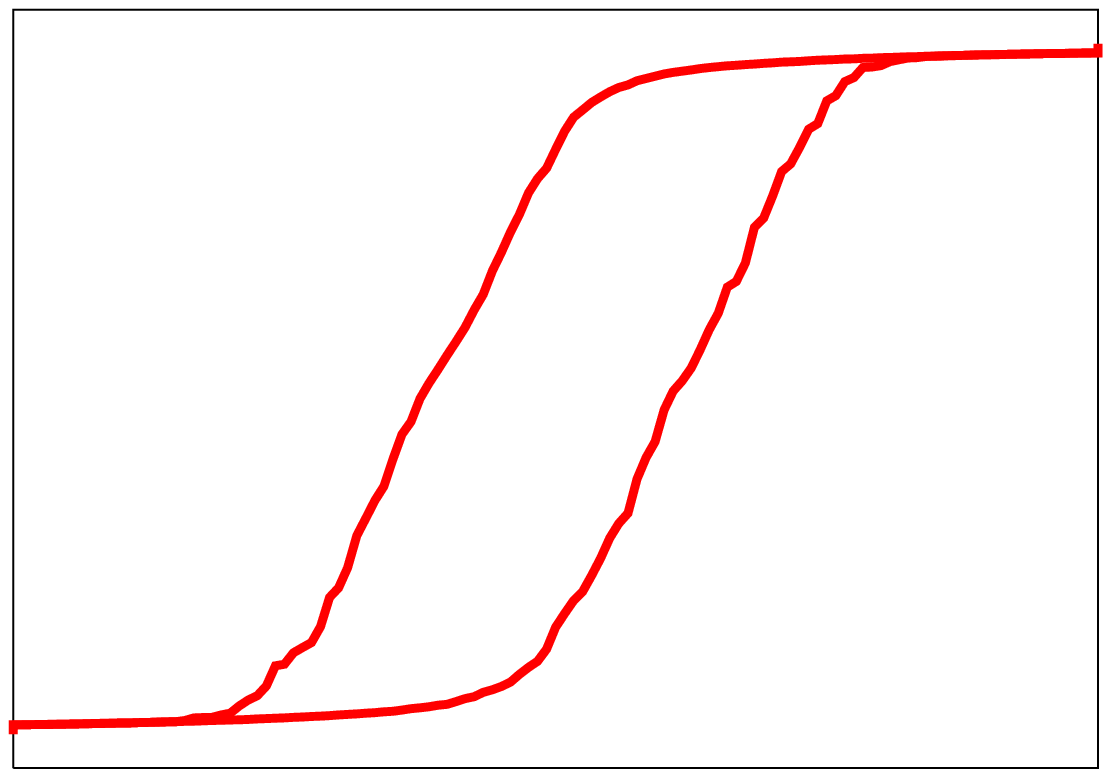}
\caption{Hysteresis loops for systems with different amounts of
disorder; the vertical axes are the magnetizations and the horizontal
axes are the external fields.  The panels correspond to the same
system as in figure \ref{disorder}.  The cliff in the hysteresis
curve vanishes as disorder is increased. }
\label{hysts}
\end{center}
\end{figure}

Disorder causes the domains to have a rougher boundary.  This rough
edge is due to the more spherical symmetric easy axes of the spins
at and near the edge.  Furthermore, the highly irregular easy axes
landscape prevents long snake-like domains from forming.  Therefore,
domains cannot grow significantly unless the external field is
lowered.  The stunting of the domain growth can be seen by a lack
of any large avalanches seen in the hysteresis loop.

Figure \ref{disorder} shows configurations near the coercive field
for systems with increasing disorder.  Figure \ref{hysts} shows the
corresponding hysteresis loops for these systems.  The configurations
and hysteresis plots are for systems with $\lambda = 1000, 11, 4.1,
3$ and $w = 0.15, 0.105, 0.08, 0.06$ starting from the upper left
to the lower right panels.  This sequence of hysteresis plots shows
how the height and slope of the cliff decreases as disorder increases.
At an intermediate amount of disorder, snakes do grow but not to
the extent of the entire system size.  The domain growth is pinned
by the disorder and growth cannot occur until the field is lowered.
Eventually, beyond $\lambda = 4.1$ and $w = 0.08$, there is no
noticeable cliff in the hysteresis curve.  A system with a lower
$w$ remains saturated for a larger range of external field, therefore
the remanent magnetization and the width of the hysteresis loop
increases as $w$ decreases.

\begin{figure}
\begin{center}
\includegraphics[width=3.25in]{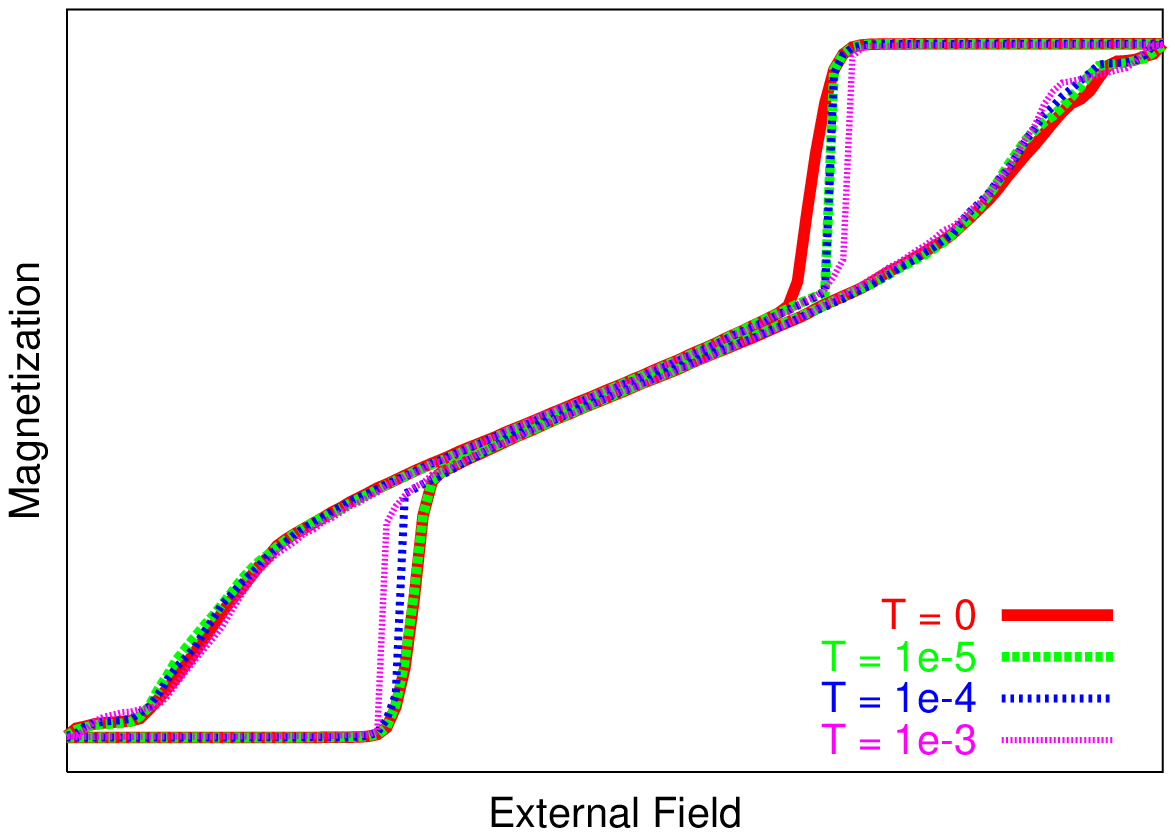} \\
\includegraphics[width=3.25in]{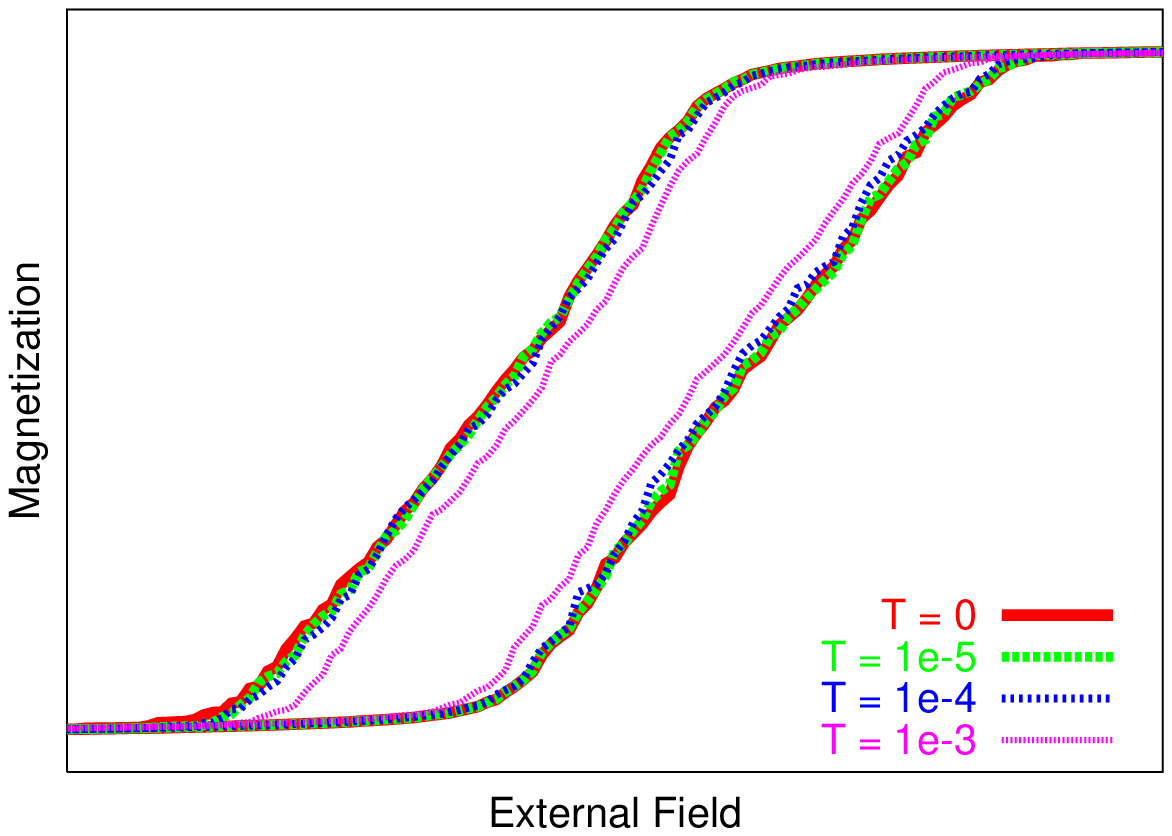}
\caption{Hysteresis loops at different temperatures for a low
disorder system (top, $\lambda = 1000, w = 0.15$) and a high disorder
system (bottom, $\lambda = 4.1, w = 0.08$).  For both systems,
thermal noise narrows the hystersis loop.  For the low disorder
system, on the descending leg thermal noise causes the cliff to
occur at a higher field.}
\label{hyst_T}
\end{center}
\end{figure}

The general features of the domain patterns and hysteresis loops
described above are dependent upon the amount of disorder but are
independent of temperature for a wide range of temperature.  However,
thermal noise does affect the field in which the domains initially
flip.  This effectively narrows the hysteresis loops as shown in
figure \ref{hyst_T}.  Another study of Co-Pt multilayers shows this
effect as well~\cite{Mobley}. Though thermal noise does have an
effect on the hysteresis loop, domain creation and growth is not
predominantly nucleation.

One can understand this by examining the low disorder system.
Starting from positive saturation, at a critical field, circular
initial domains and serpentine growth occurs for all temperatures.
At zero temperature nucleation cannot occur, therefore the zero
temperature critical field is analogous to the spinodal point.  When
thermal noise is present, it can cause domains to flip before the
spinodal field is reached, thereby nucleating domains which then
grow.  Hence at finite temperature, the cliff occurs at a field
larger than the zero temperature critical field (when all spins are
initial saturated in the positive direction).  But from figure
\ref{hyst_T}, we see that this is not a strong effect.  Nucleation
only slightly changes the field in which the cliff occurs and is
not the dominant phenomenon for domain formation and growth.

The qualitative features described in this section are observed in
the experimental samples. In summary, low disorder systems have
labyrinthine domain patterns and hysteresis loops with steep cliffs
and high disorder systems have irregular patchy domains and more
``standard'' hysteresis loops.

\section{Covariance Results}
As seen in the previous section, the amount of disorder and dipole
strength dictates the type of domain patterns and hysteresis loops.
In this section, we discuss how memory properties, specifically RPM
and CPM, are determined by the amount of disorder.  Though thermal
noise did not have a large effect on the hysteresis loops and domain
pattern, memory properties are highly sensitive to temperature.

At zero temperature, the snakes seen in the domain patterns of the
low disorder systems meander due to the small differences in one
direction versus another.  These slight differences are mainly due
to the randomness in the easy axes of the spins.  Without thermal
noise, every ascending leg will be identical to the next; a complete
field cycle that saturates the magnet returns the system to exactly
the same state.  In other words, RPM is identically equal to unity.

\begin{figure}
\begin{center}
\includegraphics[width=3.25in]{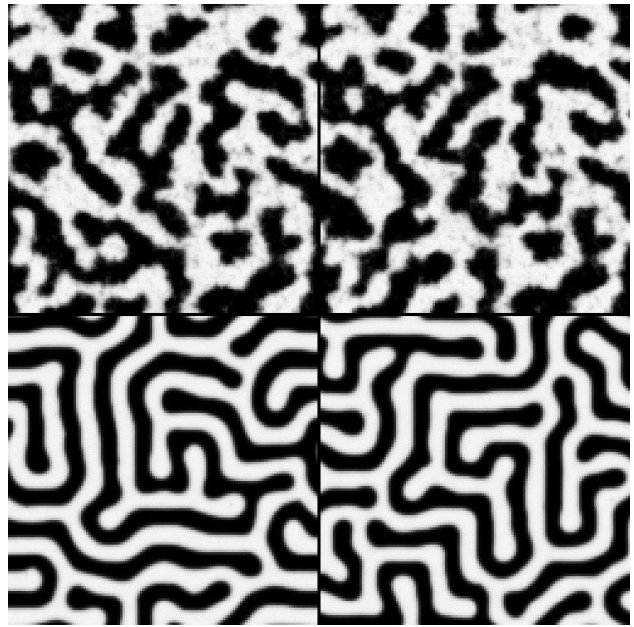}
\caption{Comparisons of the spin configurations after one complete
cycle at the coercive field. The configurations on the top row are
from a system with high disorder ($\lambda=4.1$ and $w=0.08$).
Notice the domains have essentially the same shape and are in the
same positions.  The bottom two panels are configurations from a
system with low disorder ($\lambda=1000$ and $w=0.15$). There does
not appear to be any obvious correlation between the domains for
the low disorder system.  The temperature is $10^{-5}$.}
\label{viscomp}
\end{center}
\end{figure}

Because the meandering is due to very slight differences in the
local environment of the snake, thermal noise could alter the domain
patterns dramatically.  Even when the temperature is low, the domain
configurations can be drastically different between points separated
by an integral number of complete field cycles.  In the second row
of figure \ref{viscomp}, two configurations of a low disorder system
are so different that one could not tell that both panels are the
same system separated by one field cycle.  These two configurations
both appear, at least visually, to have domains that are uncorrelated
both in their positions and their shapes.

Of course high disorder systems are also affected by thermal noise.
But because these systems are much more heterogeneous, a small
amount of noise does not cause such drastic differences.  The lack
of domain growth at constant field is due to the orientational
disorder.  This disorder also ensures that even for different
realizations of thermal noise, the domains form in essentially the
same locations with the same sizes. In other words, the relatively
more reproducible growth of domains is another consequence of pinning
by disorder. Therefore the differences in configurations after a
complete field cycle, shown in the top row of figure \ref{viscomp},
is not as drastic as was seen in the low disorder case.

Whereas thermal noise adequately explains return point differences,
explanations of the differences between complementary points also
require a discussion of the dynamics.  When the precession term of
the LLG equation is removed, the behavior of the complementary
branch is identical to the return branch (with ${\bf s} \rightarrow
-{\bf s}$ and $B_e \rightarrow -B_e$).  But when precession is
present, even at zero temperature, a configuration at field $B$
along the ascending branch is not identical (after spin reversal)
to the configuration at field $-B$ along the descending branch.

\begin{figure}
\begin{center}
\includegraphics[width=3.25in]{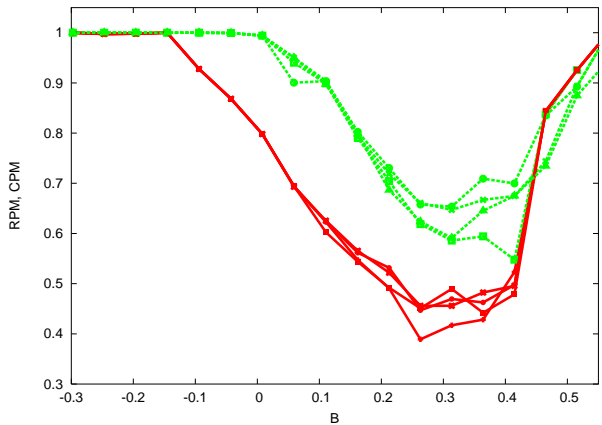}
\caption{RPM and CPM vs. external field for a $\lambda=4.1$ and
$w=0.08$ system at a very low temperature of $10^{-9}$. The covariance
values for four ``return'' legs (dashed lines) and four ``complementary''
legs (solid lines) are plotted. All ``return'' legs have essentially
the same covariance.  Similarly all ``complementary'' legs have the
same covariance. Because the system reaches saturation at the end
of every leg, RPM and CPM are independent of the number of intermediate
field cycles.}
\label{manylegs}
\end{center}
\end{figure}

As mentioned in the introduction, RPM and CPM quantify the amount
of correlation between configurations. Because each system is
saturated after every leg, RPM and CPM are independent of the number
of intermediate field cycles. At very low temperatures, the covariance
values for different legs essentially overlap.  Figure \ref{manylegs}
shows RPM and CPM plots for a highly disordered system with
$\lambda=4.1$ and $w=0.08$ at a very low temperature $T=10^{-9}$.
RPM is greater than CPM for most field points and the different RPM
legs lie very close together as do the different CPM legs.  This
feature is seen in the experiments.

\begin{figure}
\begin{center}
\includegraphics[width=3.25in]{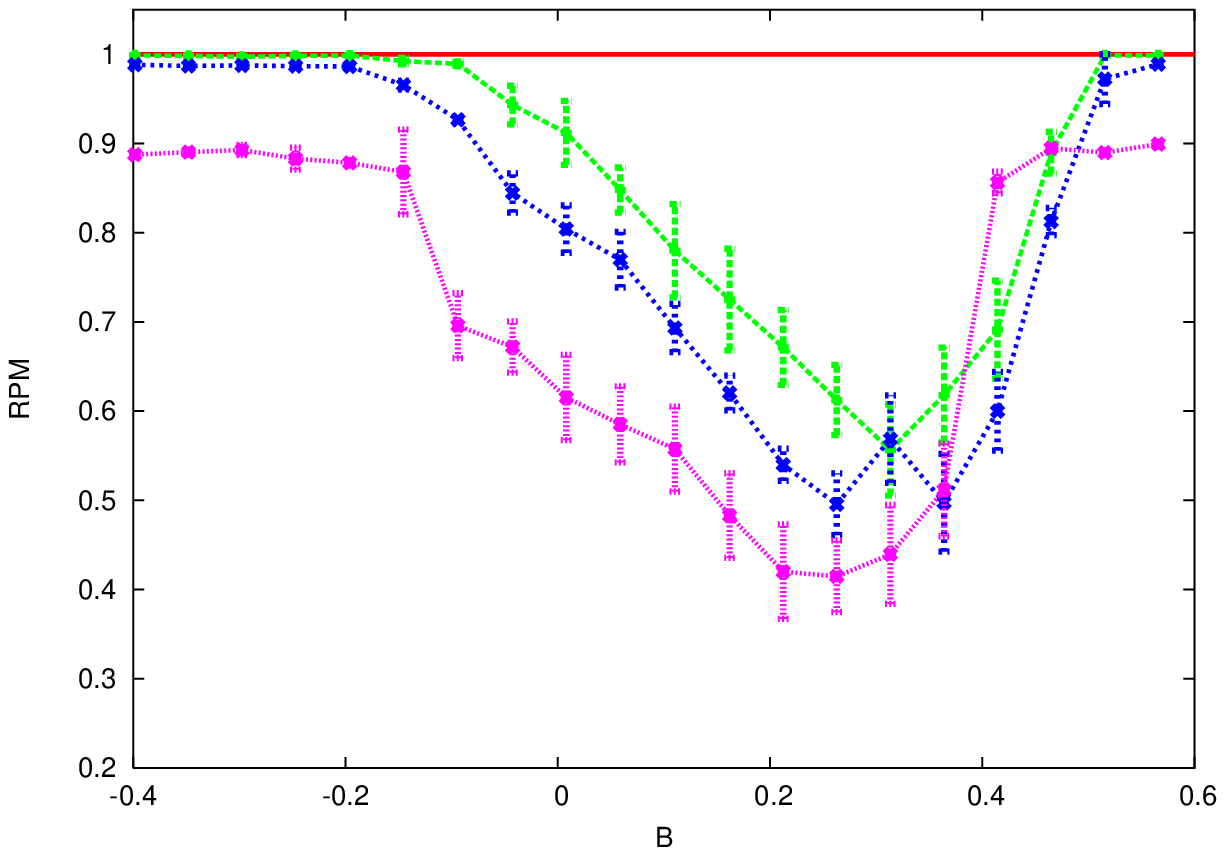} \\
\includegraphics[width=3.25in]{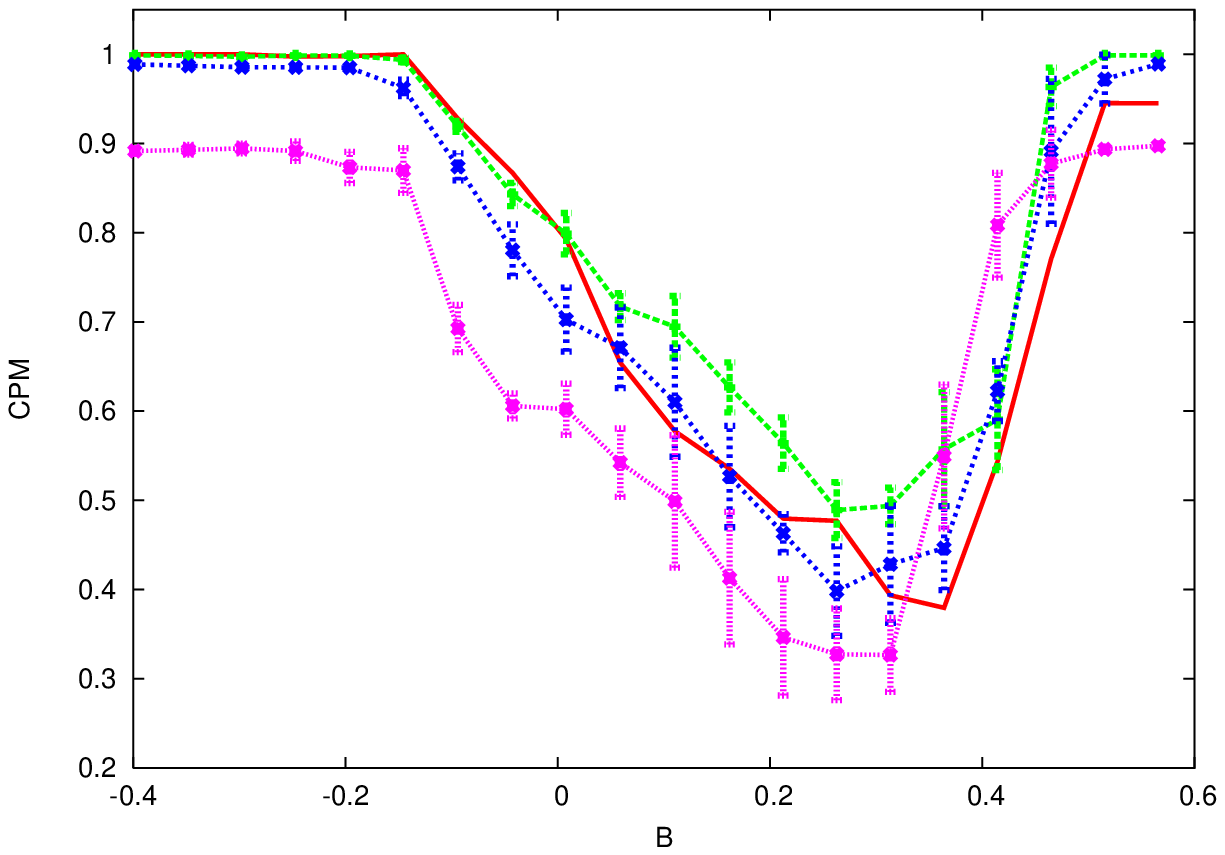}
\caption{RPM and CPM vs. external field for a highly disordered
system ($\lambda=4.1$ and $w=0.08$) at different temperatures.  At
very large negative fields, temperature is inversely related to the
height of the lines.  The temperatures are $T = 0, 10^{-5}, 10^{-4},
10^{-3}$. RPM and CPM values shown are the mean values for multiple
field cycles with RMS error bars.  The top panel shows how RPM is
exactly equal to unity for zero temperature and decreases as
temperature increases as expected.  Unexpectedly, as shown in the
bottom panel, CPM does not always decrease as temperature increases.
In fact, except for the $T=10^{-3}$ case, CPM is essentially constant
for all other temperatures within errors.  Because CPM is similar
between the zero and small temperature cases, thermal noise is not
the dominant cause of loss of complementary memory.}
\label{xpmT}
\end{center}
\end{figure}

The effects of temperature on RPM and CPM are shown for a system
with $\lambda = 4.1$ and $w = 0.08$ in figure \ref{xpmT}.  There
are much larger fluctuations compared to the system in figure
\ref{manylegs} due to the higher temperature.  As expected, these
fluctuations are reduced as the system size is increased.  The
average is done over many legs and the root mean square deviation
is shown.  These figures clearly show that at low temperature RPM
is close to unity whereas CPM is significantly lower.  As temperature
increases, RPM decreases as discussed.  Curiously, for low temperatures,
within errors, CPM stays constant.  Up to a temperature of $\sim
10^{-4}$, the dynamics appear to be dominant over temperature in
terms of complementary point memory.  This result emphasizes the
importance of the vector dynamics.  Furthermore, in the region with
RPM, CPM $< 1$ the amount of memory decreases with field.  This
decrease in memory with field is seen in the experiments as well.

\begin{figure}
\begin{center}
\includegraphics[width=1.65in]{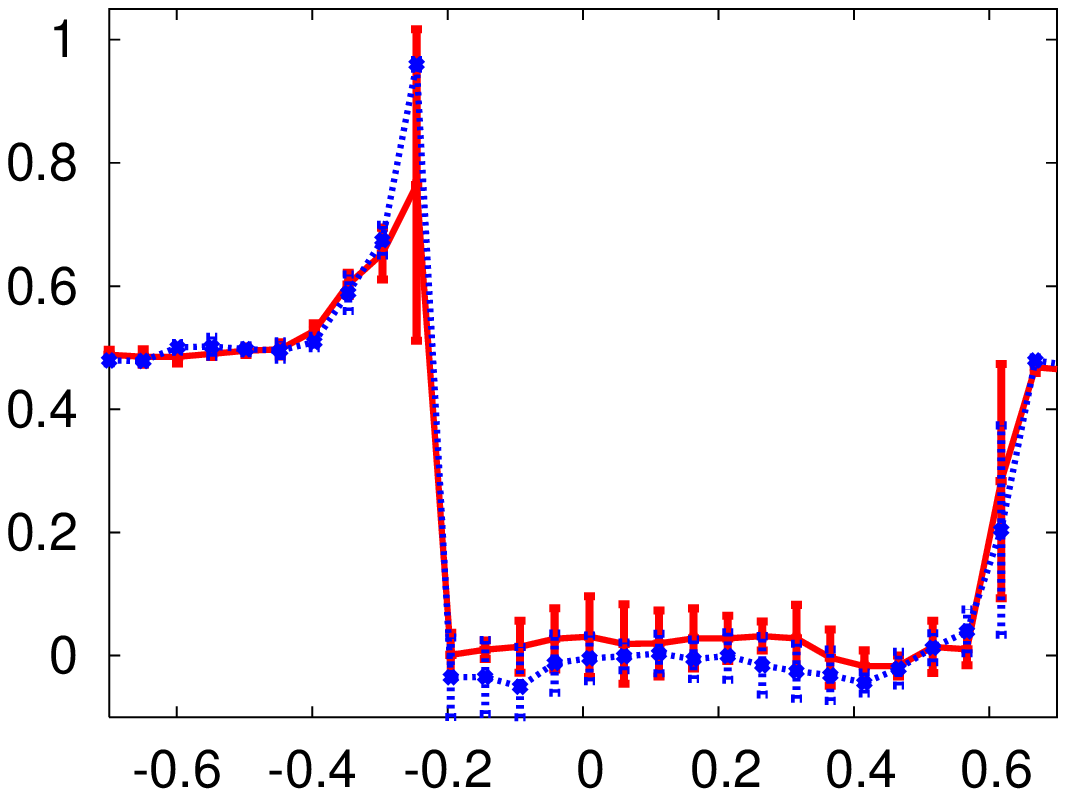}
\includegraphics[width=1.65in]{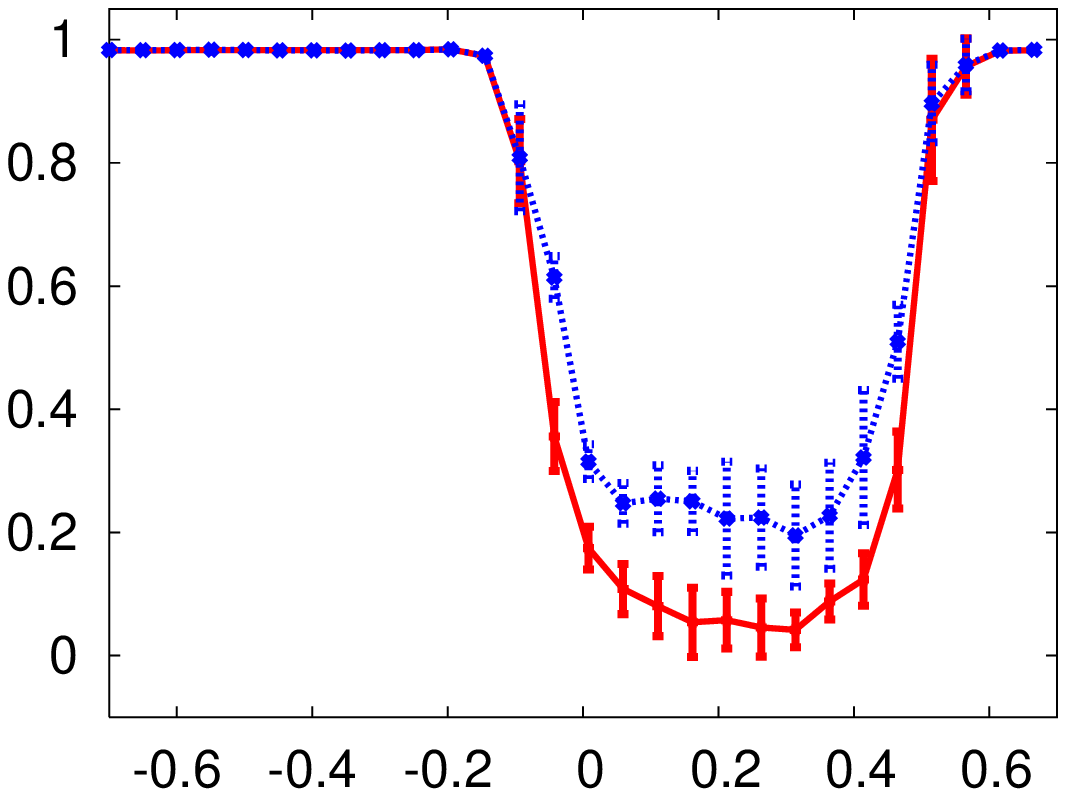}\\
\includegraphics[width=1.65in]{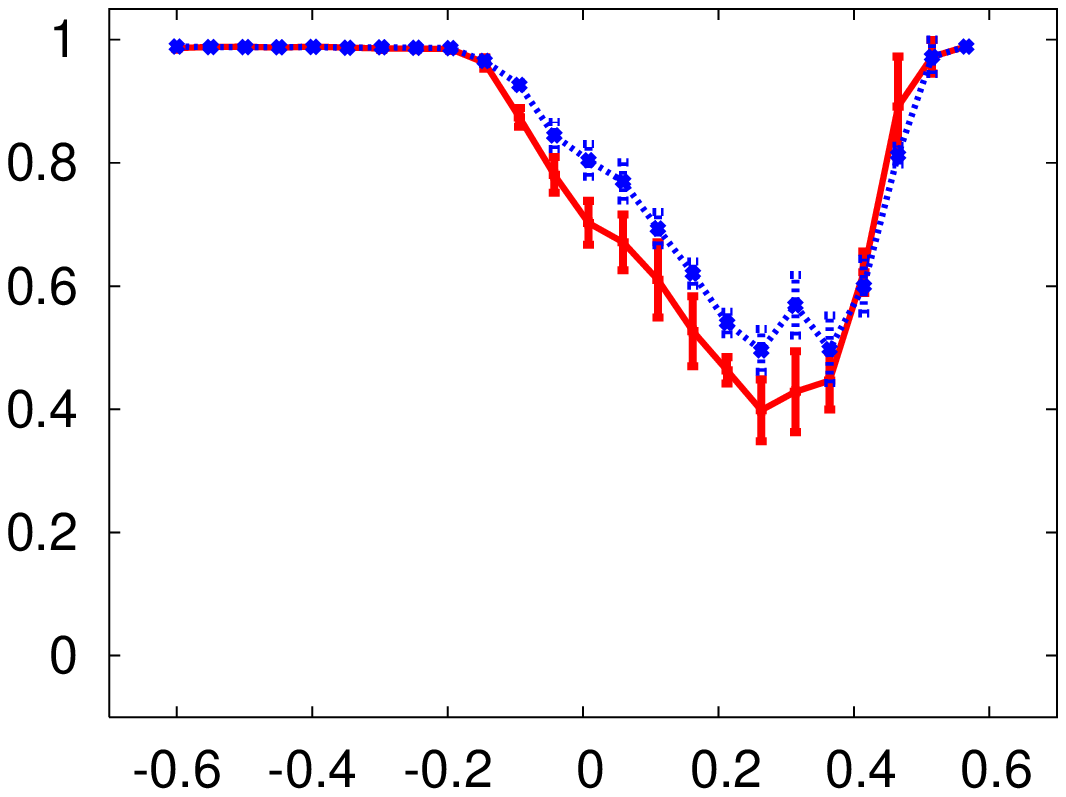}
\includegraphics[width=1.65in]{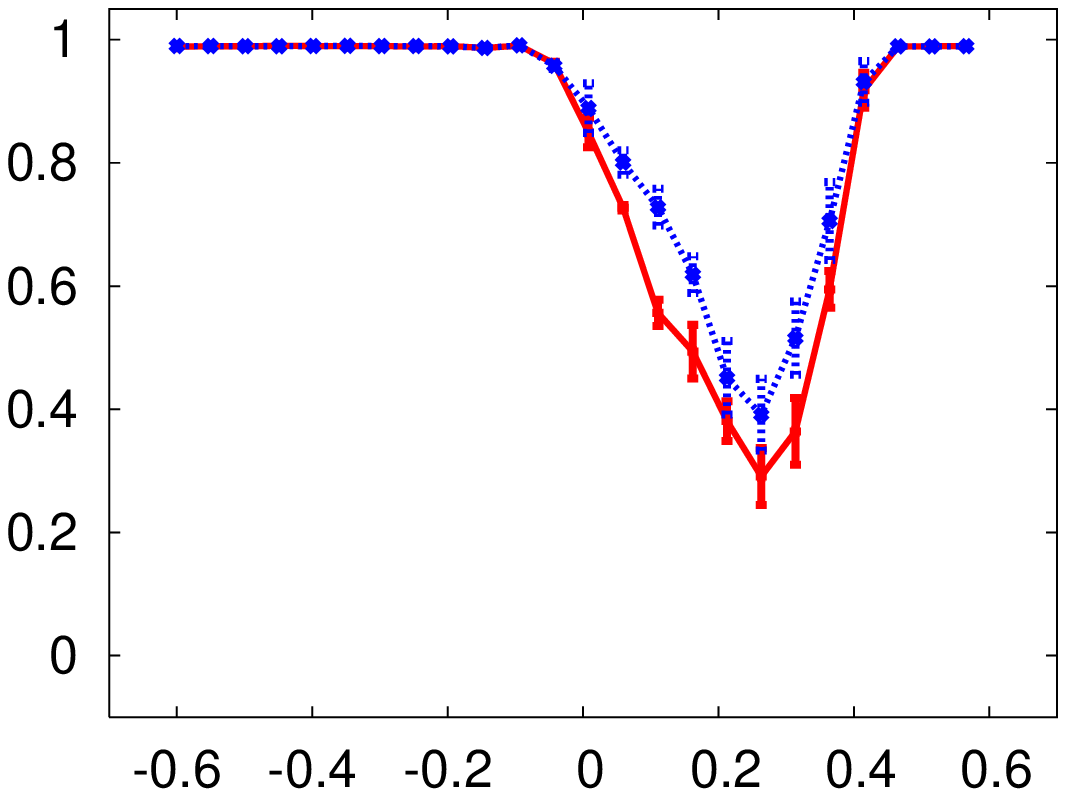}
\end{center}
\caption{RPM and CPM vs. external field as disorder is increased.
The panels here refer to the same systems as in figure \ref{disorder}
and figure \ref{hysts}.  The solid line represents CPM and the
dashed line represents RPM.  All panels have RPM$>$CPM except in
the lowest disorder system where RPM and CPM are essentially equal.
As disorder increases, both RPM and CPM increase.  The difference
between RPM and CPM is also more evident with a larger amount of
disorder.}
\label{memory}
\end{figure}

For a temperature of $10^{-4}$, figure \ref{memory} shows the average
RPM and CPM for systems with different amounts of disorder.  The
low disorder system has no significant amount of memory near the
coercive field.  Furthermore, a spike exists at the critical field
where the instabilities initially form.  This spike reveals the
fact that the initial domain(s) forms at the same position after
repeated field cycling, hence the large covariance.  RPM and CPM
drop off sharply away from this critical field due to the thermal
sensitivity of the domains.  For $\lambda \leq 4.1$ and $w \leq
0.08$ both RPM and CPM are noticeably above zero at all fields.
The general features of the RPM and CPM plots are related to the
features of the hysteresis loops as seen in figure \ref{hysts};
systems with a cliff in the hysteresis loop have much smaller RPM
and CPM than systems with more standard hysteresis loops.

\begin{figure}
\begin{center}
\includegraphics[width=3.25in]{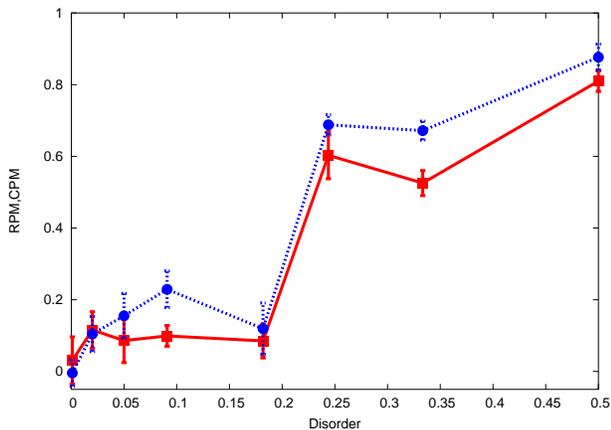}
\caption{RPM and CPM vs. ``disorder'' ($1/\lambda$) at the coercive point.  
The squares represent CPM and the circles represent RPM.  
The increase in ``memory'' with disorder is evident.}
\label{compare}
\end{center}
\end{figure}

To more clearly illustrate the relation between disorder and memory,
figure \ref{compare} contains a plot of RPM and CPM versus ``disorder''
at the coercive field. ``Disorder'' is quantified by $1/\lambda$.
There is a definite increase in memory as the amount of disorder
is increased.  Both RPM and CPM are substantial for systems with
$\lambda \gtrsim 4.1$ which corresponds to the point where the
snakes no longer appear in the domain patterns and the cliff no
longer appears in the hysteresis loop.  Similarly, the speckle
experiment found that this point is where the amount of memory is
considerable.

\section{Conclusions}
The experiments of ref.~\cite{Sorensen,Sorensen2} contain numerous
observations of the domain patterns, hysteresis loops, and memory
properties of disordered Co/Pt multilayer thin films using a variety
of techniques.  In this paper, we have attempted to understand these
results by numerically simulating the experimental systems. We have
provided a plausible explanation for all of the results found and
in particular counter-intuitive results on memory asymmetry upon
field reversal.  We have shown that vector dynamics can be crucial
in the memory property of magnetic spin systems, without which our
model would not be able to explain the experimental results.

Our simulations contain much of the qualitative features of the
domain patterns and hysteresis loops.  The domain patterns for
systems with low disorder for both experiment and simulation are
labyrinthine mazes at the coercive field.  Furthermore, we observe
snake-like growth of the domains which are responsible for the
cliffs seen in the hysteresis loops in accord with experiment.  As
the amount of disorder in the system increases, irregular patches
replace the serpentine domain patterns and the cliff in the hysteresis
loop disappears.  Because these features in the simulation resemble
the features observed in the experiments, we believe the parameters
used in the simulation are close to the actual parameters.  It would
be interesting to further investigate the growth instability in the
low disorder regime which should be similar in analysis to the
ferrofluid case~\cite{LangerGoldstein} but without conservation of
mass. It would also be interesting to see to what extent the analysis
of the dendrite problem can be carried over to this
case~\cite{langerdendrite,langerdendrite1,langersidebranches1}.

We have found that how thermal noise affects the different domain
structures determines how well the system ``remembers.'' At finite
temperature, the low disorder systems become uncorrelated much more
easily than the high disorder systems; disorder tends to pin domains,
thereby enabling the system to remember its past. The covariance
behavior of RPM as disorder is varied in the simulations agree with
much of the results from the experiments.

We are able to explain the seemingly paradoxical experimental result
that complementary points appear to ``forget'' more than return
points despite being governed by a Hamiltonian that is invariant
upon ${\bf s} \rightarrow -{\bf s}$ and ${B_e \rightarrow - B_e}$.
The non-invariance of the LLG dynamical equation provides a natural
explanation for this unexpected behavior.  Spin precession reveals
itself by decreasing the correlation between opposite legs.  Because
of the importance of precession for certain physical phenomena,
scalar theories are inadequate even for highly anisotropic materials.
From these simulation results, we show that the dynamic mechanism
is able to explain, at least qualitatively, the observations from
the speckle experiments. To make this more quantitative, a better
experimental understanding of the sputtered films is needed.

Prior to this work, Jagla~\cite{Jagla,Jagla2} produced simulations
with domain patterns and hysteresis loops remarkably similar to the
experimental ones.  In the first of his two papers~\cite{Jagla},
Jagla used a long range scalar $\phi^4$ model and obtained domain
patterns very similar to a large number of different experiments.
A $\phi^4$ theory however is not able to reproduce major hysteresis
loops correctly because $\phi$ grows indefinitely with applied
field.

In his second paper, Jagla~\cite{Jagla2} used a modified model that
did not allow the indefinite growth of $\phi$ and therefore can
produce hysteresis curves that saturate.  His model Hamiltonian
contained all the terms used here: dipolar interaction, local elastic
energy, perpendicular anisotropy, and an external field term. And
like our work, disorder was introduced in the anisotropy. However,
there are two major differences between Jagla's model and ours. The
first difference is that Jagla's model in~\cite{Jagla2} was at zero
temperature and therefore could not make predictions on the temperature
or disorder dependence of RPM and CPM. The second more important
difference is Jagla's use of a scalar model versus our vector model.
Obviously, a scalar model cannot have precession, therefore the
mechanism we propose here to explain RPM $>$ CPM cannot be applied.
And since there are no non-bilinear terms in this model (such as a
random field term), there is no mechanism for RPM to be unequal to
CPM.  However, the scalar model does produce domain patterns and
hysteresis loops similar to experiments.

Jagla also showed that when a small random field term is included
in the Hamiltonian of his original $\phi^4$ scalar model, instead
of a random anisotropy, RPM is greater than CPM~\cite{Sorensen2} and
the scalar model succeeds in explaining many of the features of the
speckle data even with a relatively small field. But an important
question remains: whether or not there exists a scalar theory that
is capable of explaining simultaneously all of the features of the
experiment that we are able to do with our vector model.

If our explanation turns out to be correct, this also has strong
implications for theory, which has often ignored the vector nature
of the dynamics and used scalar theories such as Ising models and
$\phi^4$ theories to understand these kind of systems. This opens
up the possibility that there are other unexplored consequences of
this lack of symmetry of the dynamical equation.

\begin{acknowledgments}
We thank Conor Buechler, Eric Fullerton, Olav Hellwig, Steve Kevan,
Kai Liu, Michael Pierce, and Larry Sorensen for the details and
explanations of their experiment.  We thank Eduardo Jagla, Onuttom
Narayan, and Gergely Zimanyi for very useful discussions. We also
thank Steve Kevan, Michael Pierce, and Larry Sorensen for careful
readings of the manuscript.
\end{acknowledgments}


\begin{thebibliography}{}
\bibitem{Bark}
H.Barkhausen, Z. Phys. {\bf 20}, 401 (1919).
\bibitem{Barkhaus2}
P.J. Cote and L.V. Meisel, Phys. Rev. Lett. {\bf 67} 1334 (1991).
\bibitem{Barkhaus3}
L.V. Meisel and P.J. Cote,  Phys. Rev. B {\bf 46} 10822 (1992).
\bibitem{Barkhaus4} 
B. Alessandro, C. Beatrice, G. Bertotti, and A. Montorsi, J. Appl.
Phys. {\bf 68} 2901 (1990); {\it ibid.} {\bf 68} 2908 (1990).
\bibitem{Barkhaus6}
J.S. Urbach, R.C. Madison, and J.T. Markert,  Phys. Rev. Lett.
{\bf 75} 276 (1995).
\bibitem{Barkhaus7}
O. Narayan, Phys. Rev. Lett. {\bf 77} 3855 (1996).
\bibitem{Barkhaus8}
S. Zapperi, P. Cizeau, G. Durin, and H.E. Stanley, Phys. Rev. B
{\bf 58} 6353 (1998).
\bibitem{Barkhaus9}
G. Durin and S. Zapperi, Phys. Rev. Lett. {\bf 84} 4705 (2000).
\bibitem{Sethna}
J.P. Sethna, K. Dahmen, S. Kartha, J.A. Krumhansl, B.W. Roberts and
J.D. Shore, Phys. Rev. Lett. {\bf 70}, 3347 (1993).
\bibitem{techref}
{\it Magnetic Recording Technology}, 2nd ed., edited by C. D. Mee
and E. D.  Daniel (McGraw-Hill Professional, New York 1996).
\bibitem{Sorensen}
M.S. Pierce, R.G. Moore, L.B. Sorensen, S.D., Kevan, O. Hellwig, 
E.E. Fullerton, and J.B. Kortright, Phys. Rev. Lett. {\bf 90}, 175502 
(2003).
\bibitem{Sorensen2} 
M.S. Pierce, C.R. Buechler, L.B. Sorensen, J.J. Turner, S.D. Kevan, 
E.A. Jagla, J.M. Deutsch, T. Mai, O. Narayan, J.E. Davies, K. Liu, 
J. Hunter Dunn, K.M> Chesnel, J.B. Kortright, O. Hellwig, and 
E.E. Fullerton, Phys. Rev. Lett. {\bf 94}, 017202 (2005);
and private communications.
\bibitem{Jagla}
E.A. Jagla, Phys. Rev. E {\bf 70}, 046204 (2004).
\bibitem{Jagla2}
E.A. Jagla, cond-mat/0412461.
\bibitem{pillar}
J.M. Deutsch, T. Mai and O. Narayan,  cond-mat/0408158, 
accepted for publication in Phys. Rev. E.
\bibitem{LLGref} 
F.H. de Leeuw, R. van den Doel and U. Enz, Rep. Prog.  Phys. {\bf
43}, 689 (1980).
\bibitem{LangerGoldstein}
S.A. Langer, R.E. Goldstein and D.P. Jackson, Phys. Rev. A. 
{\bf 46}, 4894 (1992).
\bibitem{Sandler}
G.M. Sandler, H.N. Bertram, T.J. Silva, and T.M. Crawford, 
J. Appl. Phys. {\bf 85}, 5080 (1999).
\bibitem{Lyberatos}
A. Lyberatos, G. Ju, R.J.M. van de Veerdonk, and D. Weller, 
J. Appl. Phys. {\bf 91}, 236 {2002}.
\bibitem{SeulMonar}
M. Seul, L.R. Monar, L. O'Gorman, and R. Wolfe, Science {\bf 254}, 
1616 (1991).
\bibitem{ferrofluid1}
A.J. Dickstein, S. Erramilli, R.E. Goldstein, D.P. Jackson, and
S.A. Langer, Science {\bf 261} 1012 (1993).
\bibitem{Mobley}
D.L. Mobley, C.R. Pike, J.E. Davies, D.L. Cox, and R.P. Singh, 
J. Phys. (Cond Mat) {\bf 16} 5897 (2004)
\bibitem{MirandaWidom}
J.A. Miranda and M. Widom, Phys. Rev. E {\bf 55} 3758 (1997).
\bibitem{langerdendrite}
J.S. Langer,  Rev. Mod. Phys. {\bf 52} 1 (1980).
\bibitem{langerdendrite1}
A. Barbieri, D.C. Hong, and J.S. Langer, Phys. Rev. A {\bf 35} 1802
(1987).
\bibitem{langersidebranches1} 
M.N. Barber, A. Barbieri, and J.S. Langer, Phys. Rev. A {\bf 36}
3340 (1987); {\it ibid.} {\bf 36} 3350 (1987).

\end{thebibliography}
\end{document}